\documentstyle[12pt,aaspp4]{article}
\newcommand{\kms}{\mbox{km\ s${^{-1}}$}}
\newcommand{\lya}{\mbox{${\rm Ly}\alpha$}}

\newcommand{\cle}{{$_ <\atop{^\sim}$}}
\newcommand{\hst}{{\sl HST}}

\newcommand{\etal}{et al.}
\newcommand{\eg}{{e.g.}}
\newcommand{\ie}{{i.e.}}
\newcommand{\halpha}{{\mbox{H$\alpha$}}}
\newcommand{\apg}{\:^{>}_{\sim}\:}
\newcommand{\apl}{\:^{<}_{\sim}\:}

\begin{document}

\title{Discovery of the Galaxy Proximity Effect and Implications for
Measurements of the Ionizing Background Radiation at Low Redshifts}

\author{Sebastian M. Pascarelle, Kenneth M. Lanzetta, \& 
Hsiao-Wen Chen\altaffilmark{1}}
\affil{Department of Physics and Astronomy, State University of New York
at Stony Brook \\ Stony Brook, NY 11794-3800, U.S.A.}
\authoremail{lanzetta@sbastr.ess.sunysb.edu}

\altaffiltext{1}{Current address: The Observatories of the Carnegie Institution
of Washington, 813 Santa Barbara Street, Pasadena, CA 91101, U.S.A.  E-Mail: 
hchen@ociw.edu}

\and

\author{John K. Webb}
\affil{School of Physics, University of New South Wales \\
Sydney, NSW 2052, AUSTRALIA}
\authoremail{jkw@bat.phys.unsw.edu.au}

\begin{abstract}

We present an analysis of galaxy and QSO absorption line pairs toward 24 QSOs 
at redshifts between $z\approx 0.2$ and 1 in an effort to establish the 
relationship between galaxies and absorption lines in physical proximity to
QSOs.  We demonstrate the existence of a {\em galaxy} proximity effect, in that
galaxies in the vicinities of QSOs do not show the same incidence and extent of
gaseous envelopes as galaxies far from QSOs.  We show that the galaxy proximity
effect exists to galaxy--QSO velocity separations of $\simeq 3000$ \kms, much
larger than the size of a typical cluster ($\simeq 1000$ \kms), {\it i.e.} it 
is more comparable to the scale of the sphere of influence of QSO ionizing
radiation rather than the scale of galaxy--QSO clustering.  This indicates that
the QSO ionizing radiation rather than some dynamical effect from the cluster 
environment is responsible for the galaxy proximity effect.  We combine 
previous findings that (1) many or most \lya\ absorption lines arise in 
extended galaxy envelopes, and (2) galaxies cluster around QSOs to show that 
the magnitude of the \lya\ forest proximity effect is underestimated. 
Consequently, determinations of the UV ionizing background intensity using the 
proximity effect are likely overestimated.  We use the galaxy--QSO 
cross-correlation function measured from our data to estimate the magnitude of
this overestimate and find that it could be as high as a factor of 20 at $z\apl
1$.  This can have strong implications for models of the origin and evolution 
of the ionizing background, and may indicate that QSOs produce sufficient 
ionizing flux at all redshifts to account for the entire background radiation 
field.

\end{abstract}


\section{Introduction}

  The forest of \lya\ absorption lines identified in the spectra of QSOs probes
physical conditions of the universe at redshifts ranging through $z \approx 5$.
Previous studies have established the redshift distribution $n(z)$ of 
\lya-forest absorption lines, with two primary results: First, $n(z)$ generally
increases with increasing redshift according to a power-law relation $n(z) 
\propto (1 + z)^\gamma$, where $\gamma$ ranges from $\gamma = 0.1$ to 0.5 at 
redshifts $z \apl 1.5$ up to $\gamma = 1.85$ to 2.78 at redshifts $z \approx 
1.5 - 4$ (Lu, Wolfe, \& Turnshek 1991, hereafter LWT91; Kulkarni \& Fall 1993,
hereafter KF93; Bechtold 1994, hereafter B94; Bahcall et al.\ 1996; Giallongo 
et al.\ 1996; Kim et al.\ 1997; Weymann et al.\ 1998; Savaglio et al.\ 1999).
Second, $n(z)$ decreases with increasing redshift along individual lines of 
sight at redshifts near the emission redshifts of the QSOs (Weymann, Carswell, 
\& Smith 1981; Carswell et al.\ 1982; Murdoch et al.\ 1986; Tytler 1987; 
Bajtlik, Duncan, \& Ostriker 1988, hereafter BDO88; LWT91; KF93; B94; Cooke,
Espey, \& Carswell 1997).  This latter effect is known as the ``proximity 
effect'' and is generally believed to exist because absorbers near QSOs (which
are subject to ionizing radiation from the background radiation field and from
the QSOs) are more highly ionized than absorbers far from QSOs (which are 
subject to ionizing radiation from only the background radiation field) and so
on average exhibit smaller neutral hydrogen column densities and weaker \lya\ 
absorption lines (BDO88).

  The primary utility of the proximity effect is that it is sensitive to the 
intensity of the background ionizing radiation field.  Under an assumption of 
ionization equilibrium, the deficit of \lya\ absorption lines (with respect to
an extrapolation of the general cosmological trend) at redshifts near the 
emission redshift of a QSO depends on the ratio of the flux of ionizing photons
from the QSO to the mean intensity of ionizing photons from the background 
ionizing radiation field.  Measurement of the proximity effect fixes this 
ratio, and measurement of the QSO spectral energy distribution fixes the flux 
of ionizing photons from the QSO, thus allowing the intensity of ionizing
photons from the background ionizing radiation field to be directly inferred. 
Results are usually expressed in terms of the mean specific intensity of the 
background ionizing radiation field at the Lyman limit, $J_{\nu_{\rm LL}}$.

  Knowledge of $J_{\nu_{\rm LL}}$ as a function of redshift is crucial for 
understanding galaxies, because ionizing radiation directly affects processes 
of star and galaxy formation. A higher level of $J_{\nu_{\rm LL}}$ at high 
redshifts would have increased the ionization fraction of the universe, 
preventing gas from collapsing down to form clouds, and consequently delaying 
or hampering star formation activity to later times ({\it e.g.}, Dekel \& Rees
1987; Efstathiou 1992).  Various groups have applied the proximity effect to 
measure $J_{\nu_{\rm LL}}$ at both low and high redshifts.  At low redshifts 
($z \apl 1$), the density of \lya-forest absorption lines is relatively low, 
and only a single (tentative) measurement of $J_{\nu_{\rm LL}}$ has been made,
yielding $J_{\nu_{\rm LL}} \approx 6.0 \times 10^{-24}$ erg s$^{-1}$ cm$^{-2}$
sr$^{-1}$ Hz$^{-1}$ (KF93).  At high redshifts ($z \approx 2 - 4$), the density
of \lya-forest absorption lines is relatively high, and various measurements of
$J_{\nu_{\rm LL}}$ in the range $J_{\nu_{\rm LL}} = 0.6 - 3.0 \times 10^{-21}$
erg s$^{-1}$ cm$^{-2}$ sr$^{-1}$ Hz$^{-1}$ have been obtained (BDO88; LWT91; 
B94; Giallongo et al.\ 1996e).  (For comparison, measurements of $J_{\nu_{\rm 
LL}}$ from observations of \halpha\ emission of Galactic and intergalactic 
clouds and photoionization edges of the neutral hydrogen disks of nearby
galaxies ({\it e.g.}, Maloney 1993) typically determine $J_{\nu_{\rm LL}}$ in 
the range $J_{\nu_{\rm LL}} \approx 1 - 10 \times 10^{-23}$ erg s$^{-1}$
cm$^{-2}$ sr$^{-1}$ Hz$^{-1}$ at redshift $z \sim 0$.)

  But previous measurements of $J_{\nu_{\rm LL}}$ from the proximity effect 
have assumed that the clouds that produce the \lya\ forest are uniformly 
distributed in space, even in the vicinities of QSOs.  This assumption may not
be valid in light of the following two points: First, direct comparison of 
galaxy and absorber redshifts along common lines of sight indicates that at 
least 50\%---and possibly all---of \lya\ absorbers at redshifts $z \apl 1$ are
associated with galaxies (Lanzetta et al.\ 1995, hereafter L95; Chen et al.\ 
1998, hereafter C98).  Second, various observations show that galaxies cluster
around QSOs (Bahcall, Schmidt, \& Gunn 1969; Hartwick \& Schade 1990; Bahcall
\& Chokshi 1991; Fisher \etal 1996; Yee \& Green 1987; Boyle, Shanks, \& Yee 
1988).  If many or most \lya\ absorbers are associated with galaxies, and if 
galaxies cluster around QSOs, then previous measurements have {\em 
underestimated} the magnitude of the proximity effect, or, equivalently, {\em 
overestimated} $J_{\nu_{\rm LL}}$ (see also Loeb \& Eisinstein 1995).  This may
have important implications for our understanding of galaxies and of the 
sources responsible for the background ionizing radiation field.

  To address these issues, we use a sample of galaxies and absorbers toward 24
low-redshift ($z \apl 1$) QSOs to examine the incidence and extent of gas 
envelopes around galaxies in the vicinities of QSOs and far from QSOs.  We find
several main results as follows: First, there exists a {\em galaxy proximity 
effect} in that galaxies in the vicinities of QSOs do not exhibit the same 
incidence and extent of gaseous envelopes as galaxies far from QSOs.  Second, 
the galaxy proximity effect appears to extend to velocity separations from the
QSOs of up to $\approx 3000$ \kms, rather than up to only $\approx 1000$ \kms, 
which suggests that the effect is related to ionizing radiation from the QSOs 
rather than to the physical environments of the QSOs.  Third, the amplitude of 
the galaxy--QSO cross-correlation function implies that previous measurements 
have overestimated $J_{\nu_{\rm LL}}$ by a factor of $\approx 20$.  While a 
standard Friedmann cosmology with $\Omega = 1$ is assumed throughout, it should
be noted that the entire analysis presented in this paper is completely
independent of the assumed value of $H_0$.

  In \S\ 2 we present the data set used in our analysis.  In \S\ 3 we 
demonstrate the existence of a galaxy proximity effect from our galaxy-absorber
pairs and discuss its origin.  In \S\ 4 we calculate the magnitude of the 
overestimate of the ionizing background radiation intensity $J_{\nu_{\rm LL}}$
derived from the proximity effect, using clustering information derived from 
the galaxy--QSO cross-correlation function.  Finally, we discuss these results 
in the context of our current understanding of the origin and evolution of 
$J_{\nu_{\rm LL}}$ in \S\ 5, and describe other effects that also can cause an
overestimate.

\section{Data}

\subsection{Galaxies}

  Observations of galaxies used in the analysis were obtained from a portion of
our imaging and spectroscopic survey of faint galaxies in the fields of Hubble
Space Telescope (\hst) spectroscopic target QSOs (L95; Lanzetta, Webb, \& 
Barcons 1996; C98).  The goal of the survey is to establish the relationship 
between galaxies and \lya\ absorption systems by directly comparing galaxies 
and \lya\ absorbers along common lines of sight.  The survey includes 523 
galaxies of magnitude $r \apl 22.5$ drawn from within a few arcmin of 24 
background QSOs.  Redshifts of the galaxies were determined from various 
ground-based spectroscopic observations, the details of which have been 
presented elsewhere (L95; Lanzetta, Webb, \& Barcons 1996; C98).  The galaxy
redshifts range from $z = 0.02$ to $\sim 1.5$, and the galaxy impact parameters
to the QSO lines of sight range from $\rho = 11$ to $1430 \ h^{-1}$ kpc.

\subsection{Absorbers}

  Observations of absorbers used in the analysis were obtained from the \hst\ 
archive.  Specifically, we analyzed Faint Object Spectrograph (FOS) spectra of 
24 QSOs of emission redshift $z_{\em} = 0.20$ to 1.07.  First, we searched the 
QSO spectra for \lya\ absorption lines according to a $5 \sigma$ detection 
threshold criterion.  Next, we searched the QSO spectra for additional \lya\
absorption lines from each of the galaxies of the sample according to a $3 
\sigma$ detection threshold criterion (L95).  We included upper limits of 
equivalent width $W< 0.35$ \AA\ for galaxies for which no \lya\ absorption 
lines were measured in the QSO spectra.  Using these criteria, we identified a 
total of 229 \lya\ absorption lines in the spectra of the 24 QSOs.  All 
absorbers included in our study have a neutral hydrogen column density $N({\rm 
H\,I}) > 10^{14}$ cm$^{-2}$.  It has been found that more than 75\% of these 
are contaminated by C\,IV (Songaila \& Cowie 1996), and such \lya\ absorbers 
are found to be strongly clustered (Fernandez-Soto \etal 1996).  Therefore, it 
follows that these absorbers are most likely to arise in galaxies at high 
redshifts.

  The equivalent width sensitivities of the spectra vary significantly, both 
from spectrum to spectrum and within individual spectra.  To account for these
variations, we measured the equivalent width sensitivity versus wavelength of 
all QSOs in the sample, based on polynomial fits to the spectra and noise 
spectra.  In this way, we employed a uniform equivalent width sensitivity 
criterion across the spectra.

\subsection{Galaxy and Absorber Pairs}

  A total of 258 of the galaxies of the sample either (1) produce a detected 
\lya\ absorption line or (2) do not produce an absorption line to within a $3 
\sigma$ limiting equivalent width threshold of 0.35 \AA.  The remaining 
galaxies have redshifts which either are larger than the redshift of the 
background QSO or would place the \lya\ absorption line in a wavelength region 
of the QSO spectrum with an equivalent width sensitivity above the $3 \sigma$ 
limiting threshold criterion.

  We established galaxy and absorber pairs using the galaxy--absorber
cross-correlation function $\xi_{ga}(\Delta v, \rho)$ measured by Lanzetta 
\etal (1997, 2001) on the basis of 3125 galaxy and absorber pairs.  
Specifically, following Chen et al.\ 1998 we formed galaxy and absorber pairs 
by requiring that (1) $\xi_{ga}(\Delta v, \rho) > 1$ and (2) $\rho < 200$ 
$h^{-1}$ kpc.  If more than one galaxy satisfied these criteria for a given 
absorption system, we ascribed the galaxy with the smallest impact parameter to
form the pair.  In total, we identified 73 galaxy-absorber pairs and 151 
galaxies that do not produce \lya\ absorption to within sensitive upper limits.
Detailed information about each of the fields studied, including coordinates of
the QSOs, emission redshifts, and spectral indices, and numbers of absorbers 
and galaxies in the field is given in Table 1.

\section{The Galaxy Proximity Effect}

  Previous work by Lanzetta \etal (1995) and Chen \etal (1998) has demonstrated
a distinct anticorrelation between \lya\ absorption line equivalent width $W$ 
and galaxy impact parameter $\rho$ at redshifts $z \apl 1$ for intervening 
galaxies far from the vicinities of background QSOs (with a line-of-sight 
velocity difference $\Delta v \apg 3000$ \kms).  In simplest terms, galaxies at
impact parameters $\rho \apl 180 \ h^{-1}$ kpc are {\em almost always} 
associated with corresponding \lya\ absorption lines, while galaxies at larger 
impact parameters rarely are.  This result is shown in the top left panel of 
Figure 1, which plots the logarithm of $W$ versus the logarithm of $\rho$.  
(Figure 1 contains data from an additional 14 QSO fields not included in the 
earlier studies, bringing the total number of fields studied to 24.)  On the 
basis of 66 galaxy and absorber pairs and 91 galaxies that do not produce 
corresponding \lya\ absorption to within sensitive upper limits, we find 
according to the generalized Kendall test that \lya\ absorption equivalent 
width is anti-correlated with galaxy impact parameter at the $7 \sigma$ level 
of significance.  Specifically, in the top left panel of Figure 1 there are 57 
galaxies with impact parameters $\rho \le 180 \ h^{-1}$ kpc, and 45 of these 
(79\%) are associated with corresponding \lya\ absorption lines, while there
are 100 galaxies with impact parameters $\rho > 180 \ h^{-1}$ kpc, and only 21 
of these (21\%) are associated with detectable absorption lines to within 
sensitive upper limits.

  The sample of galaxies discussed here and presented in the top left panel of
Figure 1 are sufficiently displaced from the background QSOs that they lack any
physical association with the QSOs, {\it i.e.}, there are no dynamical or 
radiative processes arising from the QSOs that directly affect the galaxies.  
But how do the absorption properties differ for galaxies in physical proximity 
to the background QSOs?  Can the tenuous galaxy envelopes survive the intense 
ionizing radiation from the QSO and/or the dynamical effects of the QSO cluster
environment?  If galaxies near QSOs lack the same absorption properties as 
galaxies far from QSOs, one has to distinguish between dynamical effects such 
as tidal stripping, which might exist in the cluster environment, and radiation
effects, such as the increased amount of ionizing radiation from the QSOs 
relative to the background radiation field. In an effort to make this 
distinction, we have created four subsamples from our data set in order to 
examine the incidence and extent of the gaseous envelopes of galaxies at 
various velocity separations from the background QSOs. These subsamples are (1)
the ``near'' subsample, consisting of galaxies near enough to the QSOs to be 
affected by the ionizing radiation of the QSOs, (2) the ``far'' subsample, 
presented in the top left panel of Figure 1, consisting of galaxies far enough 
from the QSOs to lie outside the range of the ionizing radiation of the QSOs, 
(3) the ``cluster'' subsample, consisting of galaxies within what we define as 
the cluster environment, and (4) the ``intermediate'' subsample, consisting of
galaxies outside the cluster environment but still subject to effects of the 
ionizing radiation of the QSOs.  These are described in detail in the following
sections.

\subsection{Extent of QSO Ionizing Radiation Field}

  The control sample for this analysis is the sample of galaxies presented in 
the top left panel of Figure 1, which we call the ``far'' subsample. It 
contains galaxies which are sufficiently far from the background QSOs so as to 
be considered completely outside the effects of the QSO radiation. To define 
the extent of the QSO radiation field we must determine the radius at which the
QSO radiation is exactly balanced by the background radiation field.  We can 
think of the ionizing radiation emitted by the QSO as forming a ``sphere of 
influence'' whose size depends on the relative strength of this radiation in 
comparison to the diffuse ionizing background radiation field.  A measure of 
this relative strength is given by the ratio of the QSO ionization rate to the 
background ionization rate,
\begin{equation}
\omega(z) = \left(\int_{\nu_{\rm LL}}^{\infty} \frac{F_\nu^Q
\sigma_\nu}{h\nu}d\nu \right) \Bigg/ \Bigg(\int_{\nu_0}^{\infty}
\frac{4\pi J_\nu \sigma_\nu}{h\nu}d\nu \Bigg) f(z)
\end{equation}
\noindent
where
\begin{equation}
f(z) = \frac{(1+z)^5}{(1+z_q)}\left[\frac{(1+z_q)^{0.5}-1}{(1+z_q)^
{0.5}-(1+z)^{0.5}}\right]^2
\end{equation}
\noindent
for $\Omega = 1$.  Here, $F_\nu^Q$ is the flux density of the QSO, $J_\nu$ is 
the specific intensity of the ionizing background radiation field, $\sigma_\nu$
is the absorption cross section of neutral hydrogen gas, $\nu_{\rm LL}$ is the
observed Lyman limit frequency at the redshift of the galaxy, and $\nu_0$ is 
the restframe Lyman limit frequency.  The cosmological factors are contained in
$f(z)$, where $z$ is the redshift of the galaxy and $z_q$ is the redshift of 
the QSO.

  To evaluate equation (1), it is first necessary to measure the spectral 
energy distributions of the QSOs and their flux densities at the Lyman limit. 
First, we created a final spectrum of each QSO by patching together the highest
signal-to-noise parts of the various FOS observations obtained through 
different gratings, which in most cases covered rest-frame wavelengths $\lambda
\approx 1000$ to 2000 \AA.  Next, we identified regions of continuum by 
selecting portions of the QSO spectra likely to be free of the many \ion{Fe}{2}
multiplets and other prominent emission features according to the composite QSO
spectrum of Francis \etal (1991).  Finally, we measured the best-fit spectral 
index $\alpha$ and Lyman limit flux density $F_0$ of each QSO by fitting a 
power-law model to the selected wavelength regions of the spectrum assuming 
$F_\nu^Q = F_0(\nu/\nu_0)^\alpha$, where $\nu_0$ is the frequency corresponding
to the Lyman limit.

  We then evaluated the integrals in equation (1), assuming first that $J_\nu$
may be approximated by $J_\nu = J_0(\nu/\nu_0)^\beta$, where $J_0 = 1.0 \times 
10^{-23}$ erg s$^{-1}$ cm$^{-2}$ sr$^{-1}$ Hz$^{-1}$ (KF93) and $\beta$ is the 
background spectral index estimated from a fit to the range of $\beta$ (for a 
background arising primarily from QSOs) given in M92 for $z=0-1$, and that the 
${\rm H\,I}$ cross section is very well approximated by $\sigma \simeq \sigma_0
(\nu/\nu_0)^{-3}$ (Osterbrock 1989).  The resulting simplified version of 
equation (1) is
\begin{equation}
\omega(z) = \frac{F_0(z)\nu_{\rm LL}^{\alpha-3}}{J_0
(z)\nu_0^{\beta-3}}\frac{\beta-3}{\alpha-3}f(z).
\end{equation}

  The extent of the QSO ionizing radiation sphere of influence, and thus the 
dividing line between the ``near'' and ``far'' subsamples, is defined to be the
point where ionization due to the QSO is exactly balanced by ionization due to 
the background radiation field, \ie, where $\omega(z)$ is unity.  We measured 
$\omega(z)$ using equation (3) for all galaxy-absorber pairs in our sample, and
Figure 2 shows a plot of log $\omega(z)$ versus log $\Delta v$ (galaxy--QSO 
velocity separation).  It can be seen from Figure 2 that, while there is an
obvious spread in log $\omega(z)$ due to the various QSO radiation intensities,
a mean value of log $\omega(z) = 0$ at log $\Delta v = 3.477$ \kms\ ($\Delta v 
= 3000$ \kms) fits quite well and different values of $J_0$ than the adopted 
$10^{-23}$ erg s$^{-1}$ cm$^{-2}$ sr$^{-1}$ Hz$^{-1}$ would have little 
consequence in the determination of subsamples.  In addition, adopting a 
different cosmological model such as LCDM would only change the calculation of 
$f(z)$ in our entire analysis.  We find that the difference in $f(z)$ between 
SCDM and LCDM would amount to no more than a factor of two at $z\apl 1$.  
Because of the small negative slope shown in Figure 2, the difference in $f(z)$
would have little effect in the determination of subsamples.  The results of 
our analysis are therefore sensitive to neither the adopted cosmological model,
nor the assumed $J_0$.

  We find that galaxies at velocity separations $\Delta v < 3000$ \kms\ from 
the background QSOs in our sample are likely to be affected by the enhanced 
ionizing radiation from the QSOs, while galaxies at greater velocity 
separations are subject only to the effects of the background ionizing 
radiation field.  Using this information, we created the ``near'' subsample of 
galaxies, which have velocity separations $\Delta v < 3000$ \kms, and the 
``far'' subsample, which have velocity separations $\Delta v > 3000$ \kms.

\subsection{Galaxy--QSO Cross-Correlation Function}

  To establish the ``cluster'' and ``intermediate'' subsamples, we must
determine a ``dynamical scale'' over which the galaxies of this study cluster 
around the QSOs of this study.  We measured the galaxy--QSO cross-correlation 
function $\xi_{gq}(\Delta v)$ in terms of galaxy--QSO velocity separation 
$\Delta v$, which is shown in Figure 3.  There is a prominent peak at $\Delta v
\simeq 0$, which indicates that the galaxies do indeed cluster around the QSOs.
We fit the cross-correlation function with a Gaussian model, which yielded 
${\rm FWHM} = 1170$ \kms.  We take galaxies to be within the cluster
environment if they occur at velocity separations at which the galaxy--QSO 
cross-correlation function exceeds unity [i.e.\ $\xi_{\rm gq}(\Delta v)>1$] in
Figure 3.  According to the Gaussian fit of $\xi_{\rm gq}(\Delta v)$, the 
clustering environment extends to $\Delta v = \pm 1180$ \kms.  Therefore, 
galaxies with $-1180 < \Delta v < 1180$ are susceptible to the dynamical 
effects of the cluster environment, and we define this subsample as the 
``cluster'' subsample.  Galaxies with $1180 < \Delta v < 3000$ \kms\ are
designated as the ``intermediate'' subsample, since they occur outside the 
clustering environment but are still susceptible to the effects of the QSO 
ionizing flux.

\subsection{Statistical Analysis}

  We defined in the previous sections four subsamples as follows: (1) the 
``near'' subsample containing galaxies with $\Delta v < 3000$ \kms, (2) the 
``far'' subsample containing galaxies with $\Delta v > 3000$ \kms, (3) the 
``cluster'' subsample containing galaxies with $|\Delta v| < 1180$ \kms, and 
(4) the ``intermediate'' subsample containing galaxies with $1180 < \Delta v <
3000$ \kms.  The next step is then to compare the $W$ (equivalent width) versus
$\rho$ (impact parameter) relation for each subsample to determine whether or 
not the absorption properties of galaxies vary among the subsamples.  A
detailed statistical analysis was carried out to perform this comparison.  We 
used Gehan (Gehan 1965) and Log-rank (Mantel 1966; Cox 1972) tests for 
incomplete observations, modified for application to multivariate observations 
(Wei \& Lachin 1984). These tests are a generalized version of the 
Kolmogorov-Smirnov test designed to properly take into account the existence of
arbitrarily censored (or failed) observations.  The purpose of these tests is 
to examine the probability of the null hypothesis, {\it i.e.}, that any two
subsamples are drawn from the same parent sample; the tests are $\chi^2$ 
distributed with respect to the null hypothesis.  In our case, the test results
will determine whether galaxies in the different subsamples share common 
physical properties of extended gas on the basis of the $W$ versus $\rho$ 
relation.  As mentioned earlier, we adopt the far subsample as the control 
sample and compare the other three subsamples.

  In making these comparisons, we must take into account the peculiar motions 
of galaxies induced by the cluster environment (Loeb \& Eisenstein 1995; also, 
{\it c.f.} Figure 3).  Some fraction of the galaxies in both the cluster 
subsample and the near subsample that do not produce absorption lines can 
actually lie behind the QSO due to these peculiar motions.  To account for 
this, we selected from our subsamples non-absorbing galaxies that lie within 
the cluster environment ($| \Delta v | < 1180$ \kms) and randomly placed them
either in front of or behind the QSO when calculating the Gehan and Log-rank 
statistics. This process was repeated 100 times, and a probability was 
calculated for the average statistics of all 100 runs.

  The results are given in Table 2 and illustrated in Figure 1, which shows $W$
versus $\rho$ for galaxies in all the subsamples.  On the basis of Figure 1 it
appears that the \lya\ absorption properties of galaxies in the near sample are
utterly unlike the \lya\ absorption properties of galaxies in the far sample.
For example, in the far sample, galaxies at impact parameters $\rho < 100 \ 
h^{-1}$ kpc are {\em almost always} associated with corresponding \lya\ 
absorption lines (30 of 32 cases), whereas in the near sample, galaxies at 
impact parameters $\rho < 100 \ h^{-1}$ kpc are {\em almost never} associated 
with corresponding \lya\ absorption lines (4 of 22 cases).  Apparently, {\em 
the only place that galaxies close to the lines of sight to background QSOs are
not associated with corresponding \lya\ absorption lines is in the immediate 
vicinities of the QSOs}.

% {\em Thus, the primary result to arise from a statistical
% analysis of the various subsamples is that there exists a
% {\em galaxy proximity effect}, in that galaxies in physical
% proximity to QSOs do not show the same incidence and extent
% of gaseous envelopes as galaxies far from QSOs.}  Furthermore,
% the statistical tests indicate that it is likely the effect
% of the QSO ionizing radiation rather than a dynamical effect
% of the cluster environment which causes the galaxy proximity
% effect.

  The statistical significance of this result is summarized in Table 2, which 
indicates that the null hypothesis (i.e.\ that the absorption properties of 
galaxies in the vicinities of QSOs are identical to the absorption properties 
of galaxies far from QSOs) can be rejected for all three of the other 
subsamples in comparison to the far subsample.  This can also be seen in the 
bottom two panels of Figure 1, which show no correlation.  Evidently, the 
anticorrelation between $W$ and $\rho$ that is seen in the far subsample does 
not exist in {\em any} of the other subsamples.  This indicates two things: (1)
there exists a {\em galaxy proximity effect}, in that galaxies in the 
vicinities of QSOs do not exhibit the same incidence and extent of gaseous 
envelopes as galaxies far from QSOs, and (2) this galaxy proximity effect is 
likely due to the increased level of ionizing radiation from the QSOs above the
mean intensity of background ionizing radiation.  If the galaxy proximity
effect instead arose from some dynamical process induced by the cluster 
environment in which galaxies were stripped of their \ion{H}{1} envelopes ({\it
e.g.}, Morris \etal 1993), then there should be a non-negligible probability of
the null hypothesis for the intermediate subsample, since these galaxies are 
outside the cluster environment.  Instead, Table 2 shows a vanishingly small 
probability for all the subsamples, so that the galaxy proximity effect 
probably exists at all velocity separations out to $\Delta v \simeq 3000$ \kms,
well outside the cluster environment. 

  We repeated the statistical tests for subsamples in which the dividing line 
between the intermediate and far subsamples was varied slightly (from $\Delta 
v = 3000$ \kms\ up to $\Delta v = 4200$ \kms to account for the effect of 
different cosmological models), and for samples in which the clustering size 
was varied slightly (from $\Delta v = \pm 600$ \kms\ up to $\Delta v = \pm 
2000$ \kms) and found that the main results are insensitive to small variations
around the values determined in this paper.

\section{Effects of Galaxy Clustering on the Proximity Effect}

Results of the previous section suggest that the deficit of absorption
lines seen in close proximity to the background QSO (the proximity
effect as discussed in \S 1) has been {\it underestimated} since the
existence of an excess of galaxies near the QSO has been neglected.
Therefore, a determination of the mean intensity of the ionizing
background radiation field $J_{\nu_{\rm LL}}$ from the proximity
effect would be {\it overestimated}. This implies that all previous
estimates of $J_{\nu_{\rm LL}}$ at various redshifts using the
proximity effect may have also been overestimated. To determine the
magnitude of this overestimate from our data set, we start by
following the definition of the proximity effect presented in BDO88.

We first model the observed number of absorption lines per unit
redshift with equivalent width above a fixed threshold as a power law
of the form
\begin{equation}
n(z) = A_0(1 + z)^\gamma
\end{equation}
\noindent
where $A_0$ is a normalization constant and $\gamma$ has been
found to range from \cle 0.3 at redshifts $<$ 1.7 (\eg, Weymann \etal
1998) to $\sim 2.0$ at redshifts between $\sim 2-4$ (\eg, B94).

Next, we describe the \ion{H}{1} column density $N_{\rm H\;{\small
I}}$ for a highly ionized absorber near a QSO as
\begin{equation}
N_{\rm H\;{\small I}} = N_0[1 + \omega(z)]^{-1}
\end{equation}
\noindent
where $N_0$ is the column density of the absorber in the absence
of the QSO and $\omega(z)$ is defined in \S 3.1.  The observed
distribution of absorber \ion{H}{1} column densities is often given by
a power law of the form
\begin{equation}
% \frac{dN}{dN_{\rm H\;{\small I}}} \propto N_{\rm H\;{\small I}}^{-\eta}
f(N_{\rm H\;{\tiny\rm I}}) \propto N_{\rm H\;{\tiny I}}^{-\eta}
\end{equation}
\noindent
so that the number of absorbers with \ion{H}{1} column density above
some threshold $N_{\rm thr}$ is
\begin{equation}
n(N_{\rm H\;{\small I}} \ge N_{\rm thr}) \propto N_{\rm H\;{\small
I}}^{-\eta+1}
\end{equation}
\noindent
Many absorption system studies have found that $\eta \approx 1.7$
(\eg, Carswell \etal 1984; Atwood, Baldwin \& Carswell 1985; Rauch
\etal 1992), which then gives
\begin{equation}
n(N_{\rm H\;{\small I}} \ge N_{\rm thr}) \propto N_{\rm H\;{\small
I}}^{-0.7}
\end{equation}

%\noindent
Therefore, for a sample of absorbers limited by the observed
\ion{H}{1} column density, the distribution of absorption lines with
redshift given in equation (4) {\em including} the proximity effect is
given by
\begin{equation}
n(z) = A_0(1 + z)^\gamma[1 + \omega(z)]^{-0.7}
\end{equation}
\noindent
However, given that most of the absorption lines in our sample are
associated with galaxies and that galaxies cluster around QSOs, the
actual number of potential absorbers must be higher by some factor
$\delta_{\rm cl}$, which is the galaxy overdensity in the vicinities
of QSOs. So the corrected version of equation (9) is
\begin{equation}
n(z) = A_0(1 + z)^\gamma[1 + \omega_{\rm
cl}(z)]^{-0.7}\delta_{\rm cl}
\end{equation}
\noindent
Equating equation (9) to equation (10) and collecting terms, we get
\begin{equation}
\left[\frac{1 + \omega(z)}{1 + \omega_{\rm cl}(z)}\right]^{-0.7} =
\delta_{\rm cl}
\end{equation}
\noindent
and since the clustering scale ($\simeq$ 1180 \kms) is smaller than
the extent of the proximity effect ($\simeq$ 3000 \kms), $\omega(z)$
will always be $\gg 1$ so that
\begin{equation}
\left[\frac{\omega(z)}{\omega_{\rm cl}(z)}\right]^{-0.7} = \delta_{\rm
cl}
\end{equation}
\noindent
Thus, the amount by which the proximity effect has been underestimated
is given by
\begin{equation}
\frac{\omega_{\rm cl}(z)}{\omega(z)} = \delta_{\rm cl}^{1.4}
\end{equation}

To determine the overdensity of galaxies around the QSOs $\delta_{\rm
cl}$, we calculated the ratio of the mean number of galaxies per unit
velocity separation at $|\Delta v| < 1180$ \kms\ to the mean number of
galaxies per unit velocity separation at $3000 < \Delta v < 50,000$
\kms.  We find from this ratio that the magnitude of the excess of
galaxies in the vicinities of the QSOs is $\delta_{\rm cl} =
8.5$. According to equation (13), then, the proximity effect measured
at low redshifts from a sample similar to the one presented in this
paper may be underestimated by a factor of 20.  Consequently, the
strength of the ionizing background radiation field $J_{\nu_{\rm LL}}$
deduced from such a measurement would be overestimated by a factor of
20.

%In agreement with these results are calculations by Loeb
% \& Eisenstein (1995) which indicate that clustering
% around QSOs at high redshifts can cause the background
% to be overestimated by up to a factor of 3. This will
% be more pronounced for weaker QSOs and/or radio-loud
% QSOs, since the clustering will be stronger and the
% proximity effect will be weaker. This is discussed
% further in \S x.

\section{Discussion}

\subsection{Comparison of Local Estimates of $J_{\nu_{\rm LL}}$}

  A quantitative understanding of the intensity of the ionizing background 
radiation, $J_{\nu_{\rm LL}}$, as a function of redshift is crucial for models
of galaxy formation and evolution, because the ionizing background radiation 
directly modulates star formation at all redshifts.  Unfortunately, it is not 
possible to measure $J_{\nu_{\rm LL}}$ directly, since the Galaxy is optically 
thick to ionizing radiation below the Lyman limit.  Therefore, we must rely on
more indirect means such as measuring its effects on detectable gas outside the
Galaxy.  At high redshifts ($z\simeq 2-4$) our knowledge of $J_{\nu_{\rm LL}}$ 
comes entirely from detections of the proximity effect ({\it e.g.}, BDO88, 
LWT91, B94) and model calculations based on the observed QSO redshift 
distribution ({\it e.g.}, M92), which tend to agree that $J_{\nu_{\rm LL}}
\simeq 1 \times 10^{-21}$ erg s$^{-1}$ cm$^{-2}$ sr$^{-1}$ Hz$^{-1}$.  Due to 
the lower density of \lya\ absorption lines at low redshifts ($z$\cle 1) as 
compared to high redshifts, detection of the proximity effect at low redshifts 
is very difficult.  Consequently, only a single tentative measurement of the
ionizing background has been reported from the proximity effect at $z\sim 0.5$ 
of $J_{\nu_{\rm LL}} = 0.2-3.6 \times 10^{-23}$ erg s$^{-1}$ cm$^{-2}$ 
sr$^{-1}$ Hz$^{-1}$ (KF93).  At these redshifts, however, various other 
techniques can be used to place constraints on $J_{\nu_{\rm LL}}$ in addition 
to the proximity effect.  In the following, we discuss these various techniques
and their results to better depict the redshift evolution of $J_{\nu_{\rm LL}}$
in the context of our results.

  There exist several estimates of $J_{\nu_{\rm LL}}$ in the local universe 
using techniques other than the proximity effect.  Since many of these 
invariably rely on some set of model assumptions, they provide only upper or 
lower limits rather than specific measurements.  For example, observations of 
a sharp cutoff in the surface density of \ion{H}{1} in the outer disks of 
nearby galaxies can constrain $J_{\nu_{\rm LL}}$ (Bochkarev \& Sunyaev 1977).
Typical estimates lie in the range $J_{\nu_{\rm LL}} \sim 1-10 \times 10^{-23}$
erg s$^{-1}$ cm$^{-2}$ sr$^{-1}$ Hz$^{-1}$, with the range of values dictated 
mainly by the assumed structure of the galactic disks (Corbelli \& Salpeter 
1993; Maloney 1993; Dove \& Shull 1994). Uncertainty in the total \ion{H}{1} 
distribution at large radii also adds to the error in such measurements.

Another technique comes from
the non-detection of H$\alpha$ emission from \ion{H}{1} clouds. An
upper limit of $J_{\nu_{\rm LL}}$ \cle 2 $\times 10^{-22}$ erg
s$^{-1}$ cm$^{-2}$ sr$^{-1}$ Hz$^{-1}$ was determined from
non-detections of H$\alpha$ emission from high-velocity clouds in the
Galaxy halo (Kutyrev \& Reynolds 1989; Songaila, Bryant, \& Cowie
1989). However, these clouds may be affected by internal sources of
ionization and/or be shielded from the intergalactic radiation
field. A more reliable measurement can come instead from intergalactic
\ion{H}{1} clouds. Two independent studies arrived at similar
stringent upper limits on $J_{\nu_{\rm LL}}$ ($J_{\nu_{\rm LL}}$ \cle
8 $\times 10^{-23}$ erg s$^{-1}$ cm$^{-2}$ sr$^{-1}$ Hz$^{-1}$) from a
lack of H$\alpha$ emission from several intergalactic clouds (Donahue,
Aldering, \& Stocke 1995; Vogel \etal 1995). Also, observations of
extraplanar and outer disk H$\alpha$ emission from several spiral
galaxies place a firm upper limit of $J_{\nu_{\rm LL}}$ \cle 1 $\times
10^{-22}$ erg s$^{-1}$ cm$^{-2}$ sr$^{-1}$ Hz$^{-1}$ (Hoopes,
Walterbos, \& Rand 1999).

  Finally, there have been several determinations of $J_{\nu_{\rm LL}}$
based on the QSO luminosity function under the assumption that a
dominant fraction of the ionizing flux at low redshift is contributed
by QSOs. Typical estimates lie in the range $J_{\nu_{\rm LL}} \sim 2-8
\times 10^{-23}$ erg s$^{-1}$ cm$^{-2}$ sr$^{-1}$ Hz$^{-1}$
(Miralda-Escud\'e \& Ostriker 1990; M92; Zuo \& Phinney 1993). More
recent calculations, taking into account updated knowledge of the
observed QSO redshift distribution find $J_{\nu_{\rm LL}} = 0.8 - 2.1
\times 10^{-23}$ erg s$^{-1}$ cm$^{-2}$ sr$^{-1}$ Hz$^{-1}$ (Shull
\etal 1999).  Furthermore, a {\em lower} 
limit of $J_{\nu_{\rm LL}} \ge 1 \times 10^{-23}$ erg s$^{-1}$ cm$^{-2}$ 
sr$^{-1}$ Hz$^{-1}$ was
suggested from measurements of the column densities of \ion{Fe}{1},
\ion{Mg}{1}, \ion{Fe}{2}, and \ion{Mg}{2} in an intergalactic
\ion{H}{1} cloud, assuming a QSO-dominated ionizing background
(Tumlinson \etal 1999). Constraints on $J_{\nu_{\rm LL}}$ from models
of the QSO emissivity versus redshift, however, must make assumptions
about the local QSO luminosity function, intrinsic QSO spectral
shapes, and radiative transfer in the intergalactic medium.

If we take as an acceptable range from the above discussion
$J_{\nu_{\rm LL}} \sim 1-8 \times 10^{-23}$ erg s$^{-1}$ cm$^{-2}$
sr$^{-1}$ Hz$^{-1}$, we can see that the KF93 measurement from the
proximity effect is quite consistent with these estimates
($J_{\nu_{\rm LL}} = 0.2 - 3.6 \times 10^{-23}$ erg s$^{-1}$ cm$^{-2}$
sr$^{-1}$ Hz$^{-1}$). If we apply the factor of 20 correction
suggested by the results of this paper to the KF93 measurement, the
intensity of the ionizing background becomes $J_{\nu_{\rm LL}} =
0.1-1.8 \times 10^{-24}$ erg s$^{-1}$ cm$^{-2}$ sr$^{-1}$ Hz$^{-1}$,
almost an order of magnitude weaker than the estimate of Shull \etal (1999).
However, these authors have fairly pointed out that their error of $J_{\nu_{\rm
LL}}$ was underestimated.  Taking a more realistic error estimate, the 
discrepancy between the two measurements may be reduced to a factor of a few. 
Because a reduced $J_{\nu_{\rm LL}}$ will result in a smaller
ionization correction, the result of our analysis also implies a smaller 
estimate for $\Omega_b$.

% $J_\nu$ \cle $3.8 \times 10^{-23}$ erg s$^{-1}$ cm$^{-2}$
% sr$^{-1}$ Hz$^{-1}$ (; Bland-Hawthorn \& Maloney 1999),

\subsection{Sources of the Ionizing Background Radiation}

An accurate depiction of the redshift evolution of $J_{\nu_{\rm LL}}$
can also help to constrain the likely candidate sources of the
ionizing background radiation, which will in turn have strong
consequences on current models of structure formation in the early
universe and the collapse and cooling of low-mass objects at early
epochs. An obvious source of ionizing radiation is the QSO
population. Initially, it was thought that the observed QSO population
did not produce enough ionizing photons at high redshifts to satisfy
measurements of $J_{\nu_{\rm LL}}$ from the proximity effect, so that
there must be other sources such as massive star formation or an
undetected QSO population (Miralda-Escud\'e \& Ostriker 1990). Later
studies found that indeed QSOs could produce (barely) enough ionizing
flux at high redshifts to ionize the universe (Meiksin \& Madau 1993),
despite the observed decrease in the QSO number density at
$z>3$. Nevertheless, the discovery of metals in high redshift \lya\
clouds implies that the intergalactic medium had been contaminated by
the products of massive star formation --- another viable source of
ionizing photons. Coupled with the existence of a large population of
star-forming galaxies at $z\sim 3$ reported over the last few years,
this implies that ionizing radiation from massive star formation at
$z>3$ can make up for the fraction that QSOs lack (Madau \& Shull
1996; Madau, Haardt, \& Rees 1999).

However, the high-redshift background ionizing spectrum is dominated
by the effects of radiative transfer in a clumpy intergalactic medium,
complicating its interpretation. The low-redshift ($z$\cle 1)
spectrum, on the other hand, should be more representative of its
sources. As at higher redshifts, there have been discrepancies at low
redshifts between the estimated background flux and the QSO space
density, which steeply declines at $z<2.5$ (Madau 1992, hereafter
M92). An underestimate of the proximity effect could then bring the
intensity of the ionizing background into closer agreement with
current models of QSOs as ionizing sources at low redshift (Giallongo,
Fontana, \& Madau 1997), especially if this underestimate is as high
as a factor of 20.

\subsection{Effects of QSO Properties on the Results}

The study presented in this paper utilized a randomly selected sample
of $z$\cle 1 QSOs of various spectral indices and radio
properties. Since QSO spectral index and radio power can have an
effect on the results of this paper, we discuss each of them
here. First, it can be seen from Figure 2 that there is a spread in
the value of $\omega(z)$ for a given velocity separation, which
implies that there is a spread in QSO ionizing flux density. The
velocity separation at which the ionizing flux from the QSO is exactly
balanced by the mean intensity of the background ionizing radiation
field defines the extent of the proximity effect, and this was found
to be $\Delta v\sim 3000$ \kms\ from our data. QSOs with larger
ionizing flux densities will push the extent of the proximity effect
out to larger velocity separations, and QSOs with smaller ionizing
flux densities will have a smaller sphere of influence. Therefore, the
proximity effect for significantly stronger QSOs will be more severely
overestimated due to galaxy clustering than the proximity effect for
weaker QSOs ({\it e.g.}, Loeb \& Eisenstein 1995).

Second, while it has been noted that galaxy clustering around low
redshift QSOs appears to be independent of QSO radio power for
$z<0.6$, there appears to be a divergence in properties at higher
redshifts (Yee \& Green 1987).  Radio-loud QSOs are often located in
rich clusters, while radio-quiet QSOs exist in smaller groups or in
the outer regions of clusters. Since most of our QSO sample (15 out of
24) have $z<0.6$, it is unlikely that radio power has any effect on
our conclusions. However, for studies conducted at higher redshifts
where the clustering properties vary significantly, the factor of 20
overestimate of $J_{\nu_{\rm LL}}$ calculated in this paper must be
adjusted accordingly, since this number directly depends on clustering
amplitude.  For example, $J_{\nu_{\rm LL}}$ measured from the
proximity effect for a sample of radio-loud QSOs at some redshift
slightly higher than $z=0.6$ should be decreased by an additional
factor of $\sim 4.6$ in comparison to a measurement based on
radio-quiet QSOs at the same redshift. (Here we assume a typical group
has about 20 members and a rich cluster has Abell richness R=1, or
about 60 members, so that the correction factor would be $\sim
3^{1.4}$.)

\subsection{Additional Contributions to the Overestimate of $J_{\nu_{\rm
LL}}$}

Galaxy clustering around QSOs is not the only source of error in
estimates of $J_{\nu_{\rm LL}}$. The mean intensity of the ionizing
background radiation can also be overestimated due to uncertainties in
the QSO redshifts (B94).  If the QSO emission redshift is measured
from broad high-ionization restframe ultraviolet lines such as \lya\
or \ion{C}{4}, then velocity shifts with respect to the QSO's systemic
redshift will cause estimates of the ionizing background from the
proximity effect to be overestimated by factors of $\sim 1.9 - 2.3$
(McIntosh \etal 1999). The low-ionization broad \ion{Mg}{2} line, or
narrow emission lines such as \ion{O}{2} or \ion{O}{3}, should instead
be used for redshift determinations, since they generally are within
$\sim 50$ \kms\ of the QSO systemic velocity. This effect, combined
with the effect of galaxy clustering from our study, implies that
estimates of the ionizing background from the proximity effect may be
overestimated by as much as a factor of $\sim 40$ if the QSO redshifts
come from the broad high-ionization ultraviolet lines. Although most
QSO redshifts in our sample have emission redshifts determined from
\ion{Mg}{2} or \ion{O}{2}, there are several that rely only on \lya\
and/or \ion{C}{4}, so that this effect should be accounted for even in
our data.

  Another source of error can come from the assumed slope of the power-law fit
to the \lya\ absorber column density distribution.  In our study, we assumed 
the widely quoted value $\eta = 1.7$.  If, for example, the slope is actually 
$\eta = 1.5$, as found by Hu \etal (1995), then the overestimate of the 
ionizing background would increase from a factor of 20 to a factor of 72.  For
a very steep column density distribution ($\eta = 2.0$) on the other hand, 
there would still need to be a correction of almost a factor of ten.  Thus, 
measurements of $J_{\nu_{\rm LL}}$ are somewhat sensitive to the assumed slope 
of the fit to the absorber column density distribution.  
%However, simulations seem to show a range of $\eta$
%that are in broad agreement with observations ({\it i.e.}, 1.4 - 1.7), so
%that very steep slopes are not likely.

  The effects discussed here tend to increase the correction factor rather 
drastically, so that estimates of $J_{\nu_{\rm LL}}$ from the proximity effect
may have to be adjusted in some cases by up to a factor of nearly 100.  This 
will have serious consequences on our current understanding of the evolution of
$J_{\nu_{\rm LL}}$.  In particular, if the differences between local estimates 
of $J_{\nu_{\rm LL}}$ and estimates at high redshifts are about a factor of 
$\sim 100$ as most studies seem to show, then any combination of the effects
discussed here could have substantial impact on our understanding of the 
evolution of $J_{\nu_{\rm LL}}$ by significantly reducing these differences.

\section{Summary}

To summarize, we have used a sample of \lya\ absorption lines and
galaxy spectra from 24 low redshift ($z$\cle 1) QSO fields to show
that there exists a galaxy proximity effect. In other words, galaxies
in the vicinities of QSOs do not show the same incidence and extent of
gaseous envelopes as galaxies far from QSOs. We find that the scale of
the galaxy proximity effect is consistent with the scale of the QSO
ionizing radiation field ($\Delta v$ \cle 3000 \kms) rather than the
scale of galaxy clustering ($\Delta v$ \cle 1180 \kms) for our data,
indicating that the galaxy proximity effect is due to the increased
ionizing radiation from the QSO rather than some cluster environmental
effect. We furthermore find that since (1) most \lya\ absorption
systems arise in galaxies, and (2) galaxies cluster around QSOs, the
strength of the proximity effect has likely been underestimated. That
is to say, there are more potential \lya\ absorbers (galaxies) in the
vicinities of QSOs than assumed if clustering is ignored, leading to
an underestimate of the magnitude of the proximity
effect. Consequently, the mean intensity of the ionizing background
radiation $J_{\nu_{\rm LL}}$ as determined from the proximity effect
will be overestimated. We find the overestimate to be as high as a
factor of 20 at low redshifts (higher if other effects are taken into
account), which brings estimates of $J_{\nu_{\rm LL}}$ down to a level
which may make it easier to reconcile with models of QSOs as the
primary source of ionizing photons.

\acknowledgments

SMP, HWC, and KML acknowledge support from NASA grant NAGW-4422 and
NSF grant AST--9624216.  JKW acknowledges support from the Australian
Research Council.

\newpage

\figcaption{({\em top left}) Logarithm of rest-frame equivalent width $W$ versus
logarithm of impact parameter $\rho$ plotted for the ``far'' sample of
galaxies, i.e. those galaxies with velocity separations from the background
QSOs greater than 3000 \kms.  Open circles with arrows indicated $3 \sigma$
upper limits to $W$ of 0.35 \AA.  Note how galaxies at small impact parameters
are more likely to produce absorption lines than are galaxies at large impact
parameters, and note the marked trend from large $W$ at small $\rho$ to small
$W$ at large $\rho$. ({\em top right})  Logarithm of rest-frame equivalent
width $W$ versus logarithm of impact parameter $\rho$ for the ``near'' sample
of galaxies, i.e. those galaxies with velocity separations from the background
QSOs of less than 3000 km s$^{-1}$.  Note how most galaxies in this sample do
not produce an absorption line at {\em any} impact parameter.  ({\em bottom
left}) Logright of rest-frame equivalent width $W$ versus logarithm of impact
parameter $\rho$ for the cluster subsample of galaxies.  ({\em bottom right})
Logarithm of rest-frame equivalent width $W$ versus logarithm of impact
parameter $\rho$ for the intermediate subsample of galaxies.}

\figcaption{Log $\omega(z)$ versus log $\Delta$v for all
galaxies--absorber pairs in the data set. The point at which
the QSO ionizing radiation is exactly balanced by the
diffuse ionizing background radiation field by definition
occurs at $\omega=1$, which corresponds to a galaxy--QSO
velocity separation of $\Delta$v = 3000 km s$^{-1}$ in
this plot.}

\figcaption{The galaxy--QSO cross-correlation function versus
galaxy--QSO velocity separation. Note the prominent peak
at a velocity separation of zero, indicating that galaxies
do indeed cluster around the QSOs. A Gaussian fit to the
peak indicates that the typical cluster environment for
this sample of galaxies extends out to galaxy--QSO
velocity separations of $\approx \pm 1180$ km s$^{-1}$.}

\newpage

\begin{center}
\begin{tabular}{p{1.5in}ccccccc}
\multicolumn{8}{c}{TABLE 1---QSO SAMPLE} \\
\hline
\hline
\multicolumn{1}{c}{QSO Field} & RA (J2000) & Dec (J2000) & $z_{\rm em}$ &
$\alpha$ & $F_0$$^a$ & $n_{\rm abs}$ & $n_{\rm gxy}$ \\
\hline
0044$+$0303 \dotfill & 00$^{\rm h}$ 47$^{\rm m}$ 05\fs 91 & $+$03\arcdeg 19\arcmin
55\farcs 0 & 0.62326 & $-$1.12 & 0.55  &  6 & 5 \\
0122$-$0021 \dotfill & 01$^{\rm h}$ 25$^{\rm m}$ 28\fs 84 & $-$00\arcdeg 05\arcmin
55\farcs 9 & 1.070   & $-$1.55 & 0.54  & 20 & 3 \\
0349$-$1438 \dotfill & 03$^{\rm h}$ 51$^{\rm m}$ 28\fs 54 & $-$14\arcdeg 29\arcmin
08\farcs 7 & 0.61625 & $-$0.58 & 1.29  &  5 & 7 \\
0405$-$1219 \dotfill & 04$^{\rm h}$ 07$^{\rm m}$ 48\fs 43 & $-$12\arcdeg 11\arcmin
36\farcs 7 & 0.57259 & $-$0.71 & 2.46  & 15 & 26 \\
0454$-$2203 \dotfill & 04$^{\rm h}$ 56$^{\rm m}$ 08\fs 90 & $-$21\arcdeg 59\arcmin
09\farcs 0 & 0.53348 & $-$1.05 & 1.02  & 15 & 10 \\
0637$-$7513 \dotfill & 06$^{\rm h}$ 35$^{\rm m}$ 46\fs 51 & $-$75\arcdeg 16\arcmin
16\farcs 8 & 0.656   & $-$1.15 & 0.64  & 10 & 12 \\
0850$+$4400 \dotfill & 08$^{\rm h}$ 53$^{\rm m}$ 34\fs 20 & $+$43\arcdeg 49\arcmin
01\farcs 0 & 0.51390 & $-$0.48 & 0.37  &  1 & 5 \\
0903$+$1658 \dotfill & 09$^{\rm h}$ 06$^{\rm m}$ 31\fs 92 & $+$16\arcdeg 46\arcmin
12\farcs 8 & 0.4121  & $-$0.79 & 0.10  &  1 & 15 \\
1001$+$2910 \dotfill & 10$^{\rm h}$ 04$^{\rm m}$ 02\fs 63 & $+$28\arcdeg 55\arcmin
35\farcs 5 & 0.32970 & $-$0.36 & 1.31  & 16 & 11 \\
1049$-$0035 \dotfill & 10$^{\rm h}$ 51$^{\rm m}$ 51\fs 50 & $-$00\arcdeg 51\arcmin
16\farcs 6 & 0.35990 & $-$1.10 & 0.81  &  1 & 2 \\
1136$-$1334 \dotfill & 11$^{\rm h}$ 39$^{\rm m}$ 10\fs 70 & $-$13\arcdeg 50\arcmin
43\farcs 5 & 0.557   & $-$0.83 & 0.55  &  2 & 13 \\
1216$+$0657 \dotfill & 12$^{\rm h}$ 19$^{\rm m}$ 20\fs 88 & $+$06\arcdeg 38\arcmin
38\farcs 4 & 0.33130 & $-$0.74 & 1.31  &  8 & 11 \\
1259$+$5920 \dotfill & 13$^{\rm h}$ 01$^{\rm m}$ 12\fs 90 & $+$59\arcdeg 02\arcmin
06\farcs 4 & 0.47780 & $-$0.85 & 1.47  &  8 & 17 \\
1317$+$2743 \dotfill & 13$^{\rm h}$ 19$^{\rm m}$ 56\fs 32 & $+$27\arcdeg 28\arcmin
08\farcs 6 & 1.022   & $-$0.48 & 1.18  & 38 & 12 \\
1354$+$1933 \dotfill & 13$^{\rm h}$ 57$^{\rm m}$ 04\fs 44 & $+$19\arcdeg 19\arcmin
07\farcs 4 & 0.719   & $-$1.67 & 0.69  & 10 & 5 \\
1424$-$1150 \dotfill & 14$^{\rm h}$ 27$^{\rm m}$ 38\fs 17 & $-$12\arcdeg 03\arcmin
50\farcs 6 & 0.806   & $-$1.88 & 0.38  &  8 & 8 \\
1545$+$2101 \dotfill & 15$^{\rm h}$ 47$^{\rm m}$ 43\fs 54 & $+$20\arcdeg 52\arcmin
16\farcs 7 & 0.26430 & $-$0.91 & 0.71  &  4 & 7 \\
1622$+$2352 \dotfill & 16$^{\rm h}$ 24$^{\rm m}$ 39\fs 08 & $+$23\arcdeg 45\arcmin
12\farcs 8 & 0.927   & $-$2.23 & 0.12  & 12 & 17 \\
1641$+$3954 \dotfill & 16$^{\rm h}$ 42$^{\rm m}$ 58\fs 81 & $+$39\arcdeg 48\arcmin
37\farcs 0 & 0.59280 & $-$1.57 & 0.67  &  2 & 20 \\
1704$+$6048 \dotfill & 17$^{\rm h}$ 04$^{\rm m}$ 41\fs 35 & $+$60\arcdeg 44\arcmin
30\farcs 3 & 0.37210 & $-$0.83 & 1.40  & 13 & 25 \\
1821$+$6419 \dotfill & 18$^{\rm h}$ 21$^{\rm m}$ 34\fs 38 & $+$64\arcdeg 20\arcmin
59\farcs 6 & 0.2977  & $-$1.17 & 5.56  & 13 & 17 \\
2135$-$1446 \dotfill & 21$^{\rm h}$ 37$^{\rm m}$ 45\fs 24 & $-$14\arcdeg 32\arcmin
55\farcs 4 & 0.20030 & $-$0.23 & 0.39  &  5 & 7 \\
2141$+$1730 \dotfill & 21$^{\rm h}$ 43$^{\rm m}$ 35\fs 55 & $+$17\arcdeg 43\arcmin
49\farcs 3 & 0.21110 & $-$1.38 & 0.81  &  1 & 1 \\
2251$+$1552 \dotfill & 22$^{\rm h}$ 53$^{\rm m}$ 57\fs 75 & $+$16\arcdeg 08\arcmin
53\farcs 6 & 0.859   & $-$2.88 & 0.74  & 15 & 2 \\
\hline
\multicolumn{8}{l}{$^a$$10^{26}$ erg s$^{-1}$ cm$^{-2}$ Hz$^{-1}$.}
\\
\end{tabular}
\end{center}

\newpage

\begin{center}
\begin{tabular}{p{1.5in}ccc}
\multicolumn{4}{c}{TABLE 2---PROBABILITY OF NULL HYPOTHESIS} \\
\hline
\hline
\multicolumn{1}{c}{Sample} & $\Delta v$ (\kms) & Gehan (\%) & Log-rank (\%) \\
\hline
near \dotfill         & $< 3000$       & $6.5 \times 10^{-5}$ &
$7.1 \times 10^{-5}$ \\
intermediate \dotfill & $1180 - 3000$  & $1.2 \times 10^{-4}$ &
$3.0 \times 10^{-3}$ \\
cluster \dotfill      & $-1180 - 1180$ & $6.2 \times 10^{-4}$ &
$3.4 \times 10^{-4}$ \\
\hline
\end{tabular}
\end{center}

\newpage

\plotone{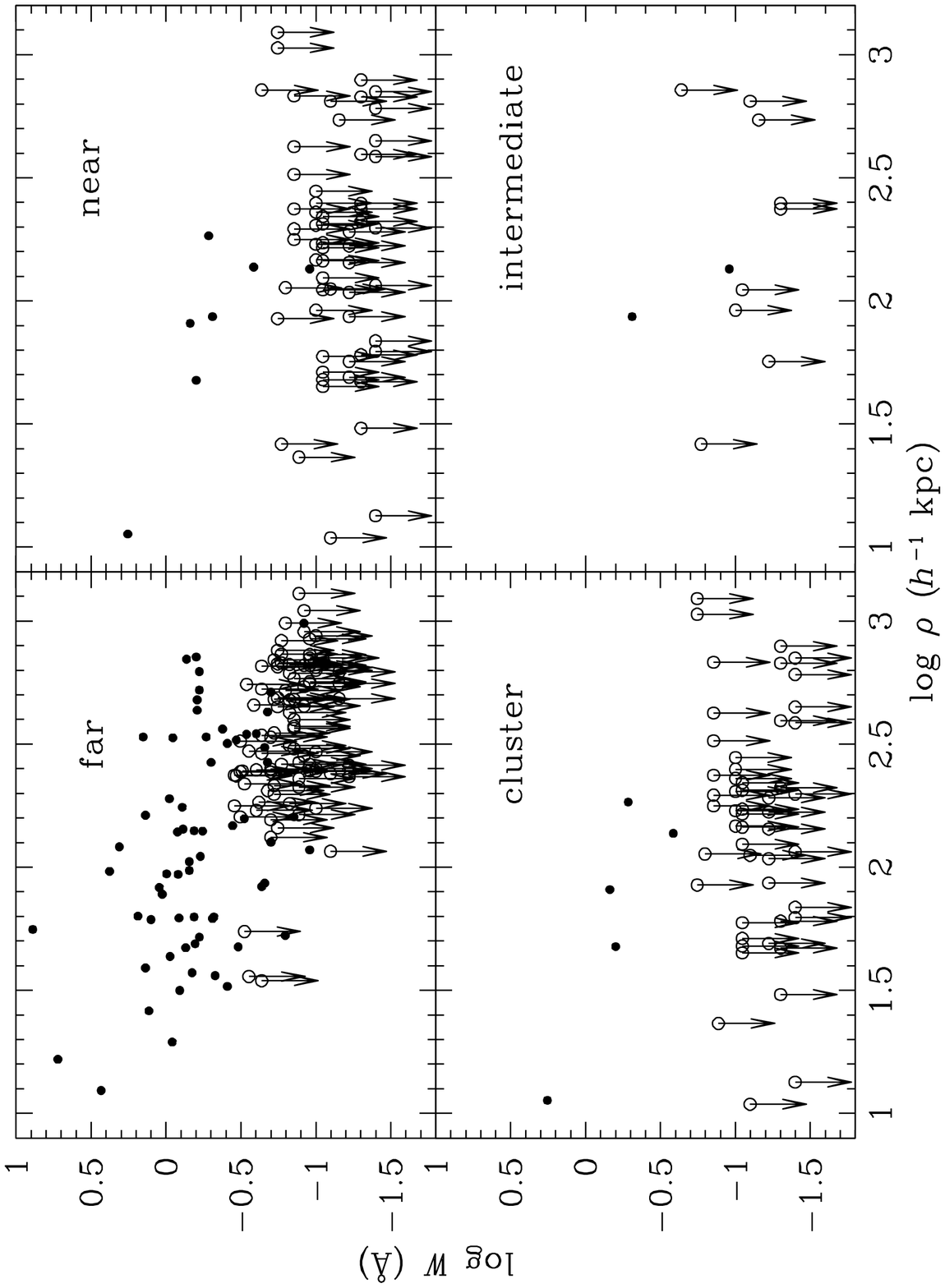}

\newpage

\plotone{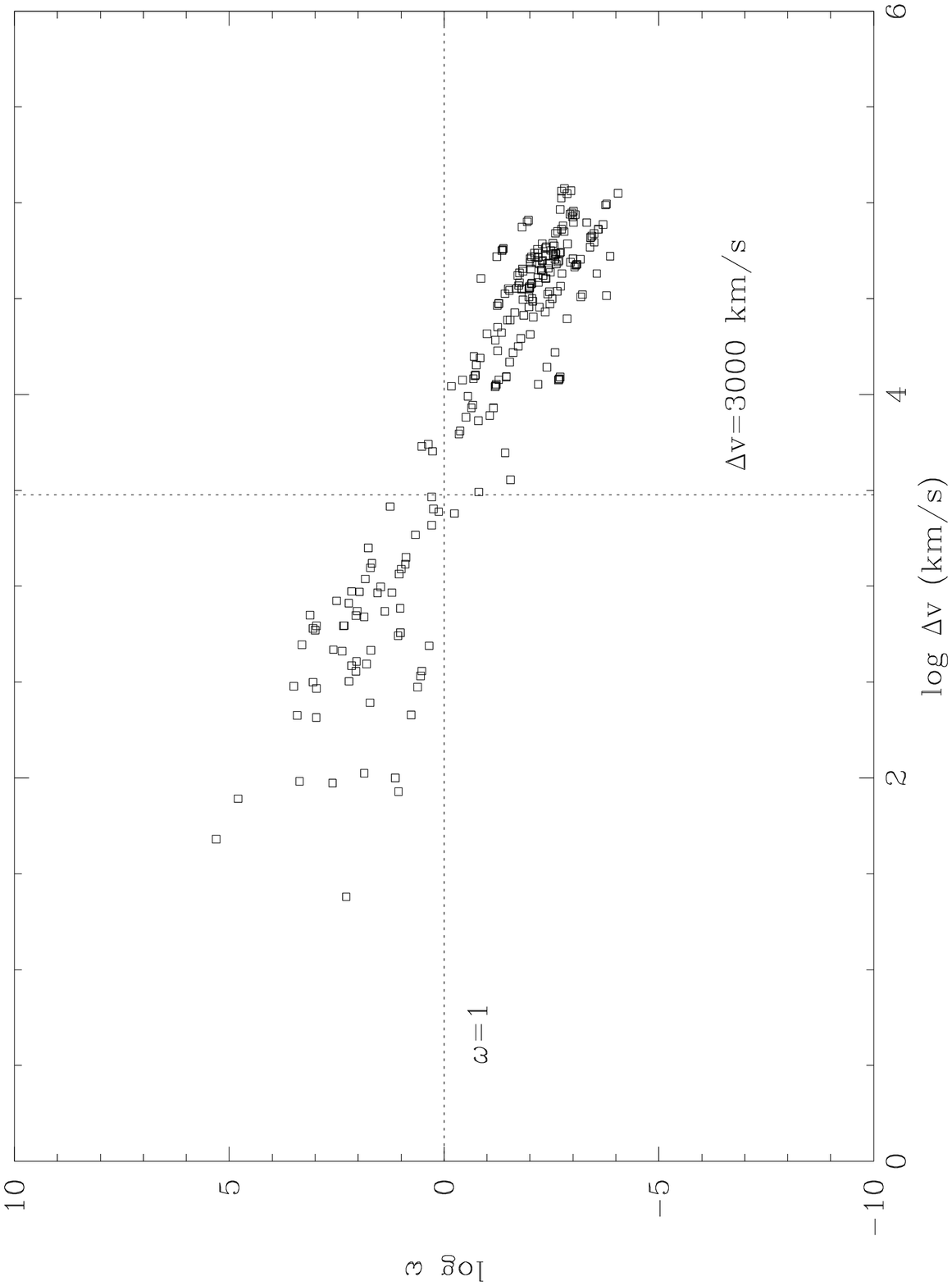}

\newpage

\plotone{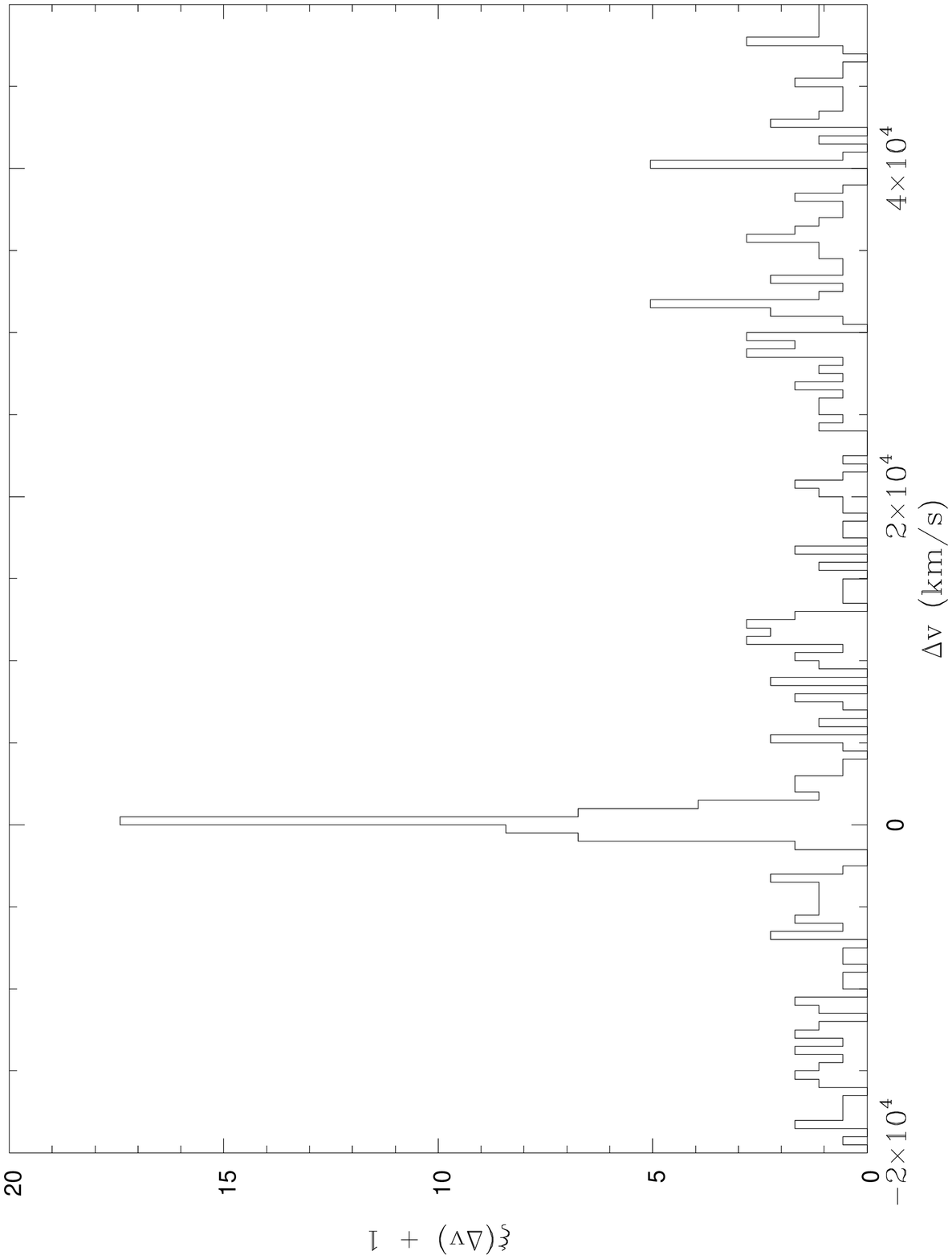}

\end{document}